\documentclass[%
 reprint,
 amsmath,amssymb,
 aps,
nofootinbib]{revtex4-1}

\usepackage{dsfont}
\usepackage{hyperref}
\usepackage{graphicx}
\usepackage{dcolumn}
\usepackage{bm}
\usepackage{amsmath}

\begin{document}


\title{An information and field theoretic approach to the grand canonical ensemble}

\author{Diego Granziol}
\email{diego@robots.ox.ac.uk}
\author{Stephen Roberts}%
\email{sjrob@robots.ox.ac.uk}
\affiliation{%
Machine Learning Research Group \& Oxford-Man Institute, University of Oxford, U.K.\\
}%

\date{\today}

\begin{abstract}
We present a novel derivation of the constraints required to obtain the underlying principles of statistical mechanics using a maximum entropy framework. We derive the mean value constraints by use of the central limit theorem and the scaling properties of Lagrange multipliers. We then arrive at the same result using a quantum free field theory and the Ward identities. The work provides a principled footing for maximum entropy methods in statistical physics, adding the body of work aligned to Jaynes's vision of statistical mechanics as a form of inference rather than a physical theory dependent on ergodicity, metric transitivity and equal a priori probabilities \cite{Jaynes1967}. We show that statistical independence, in the macroscopic limit, is the unifying concept that leads to all these derivations.
\begin{description}
\item[PACS numbers]
\pacs\verb{89.70.Cf,05.70.-a,05.20.Gg,05.30.Ch}
\end{description}
\end{abstract}

\maketitle


\section{\label{sec:level1}Introduction}
	\subsection{What is Entropy?}
	The method of maximum entropy (MaxEnt) \cite{maxentreview} is a method for generating the least biased estimate possible on the given information, it is maximally non-committal with regard to missing information \cite{inftheoryjaynes}. Mathematically we maximise the functional 
	\begin{equation}
	\label{BSG}
	S = \int p(\vec{x})\log p(\vec{x})d\vec{x}- \sum_{i}\lambda_{i}\bigg[\int p(\vec{x})f_{i}(\vec{x})d\vec{x} - \mu_{i}\bigg]
	\end{equation}
	with respect to $p(\vec{x})$, where $\langle f_{i}(\vec{x})\rangle = \mu_{i}$ are the constraints. It can be proved under the axioms of consistency, uniqueness and independence, that for constraints in the form of expected values, it is the unique functional who's extremization represents a valid chain of inference \cite{shore1980axiomatic}. It has been widely and successfully applied, from condensed matter physics \cite{giffin} to finance \cite{nerioptions,entropybuchen}. Along with its path equivalent, maximum caliber (MaxCal) \cite{Gonzalez2014}, it has been used to successfully derive statistical mechanics, non-relativistic quantum mechanics, Newton's laws and Bayes' rule \cite{Gonzalez2014,caticha2012entropic}. 

	\subsection{Where do we stand?}
	In his influential 1957 paper, introducing the concept of MaxEnt, E. T. Jaynes concludes ``that statistical mechanics need not be regarded as a physical theory dependent for its validity on the truth of additional assumptions, such as ergodicity, metric transitivity and equal a priori probabilities" \cite{inftheoryjaynes}. These are considered irrelevant artefacts of a ``dogmatic, single-minded insistence of the frequency theory of probability" \cite{Jaynes1967,brandeis}. By ``know[ing] only the average $\langle E \rangle$" or ``given the expectation values $\langle E \rangle, \langle n_{1} \rangle, \langle n_{2} \rangle$" Jaynes derives the canonical/quantum grand-canonical ensemble respectively. Yet these crucial constraints, on which the crux of the entire theory is based are seemingly introduced unsuspectingly. We find an argument based on mean square deviations being within reasonable experimental error \cite{Jaynes1967} to neglect higher order moment information. In \cite{julianandpresse}, Lee and Press\'e look to derive the uniform micro-canonical ensemble as a consequence of maximizing the Gibbs-Shannon entropy. However, in their derivations we note that the uniform distribution, which they use to describe the microcanonical ensemble, is not a valid density as 
	\begin{equation}
	    \sum_{i} p_{i} = \sum_{i} \frac{\delta(E-E_{i})}{\Omega(E)} = \frac{1}{\Omega(E)} \neq 1,
	\end{equation}
	where we note that the sum over the discrete delta is unity. However, extending the domain of the probability density over both $i$ and $E_{j}$ we have\footnote{We note that it is possible that, instead of the delta distribution, they meant the indicator function $\mathds{1}(E-E_{i})$, this is equivalent to equation \eqref{2dpeq}.} 
		\begin{equation}
		\label{2dpeq}
	    \sum_{i,E_{j}} p_{i,E_{j}} = \sum_{i,E_{j}} \frac{\delta(E_{j}-E_{i})}{\Omega(E)} = \frac{\sum_{E_{j}}}{\Omega(E)} = 1,
	 	\end{equation}
	where $\sum_{E_{j}} = \Omega(E)$. It is clear to see that the condition imposed, $\sum_{i}p_{i}(1-\delta_{E_{i}E}) = 0$, leads to this ``derived distribution" independent of the method of maximum entropy. This condition is usually justified through the use of ergodicity and metric transitivity \cite{chandler} (appendix \ref{ergodicity}). We write their $p_{i}$ as $p_{i,E_{j}}$ to denote the different states of equal energy.
	\begin{equation}
	\begin{aligned}
	& \sum_{i,E_{j}}p_{i,E_{j}}(1-\delta_{E_{i}E_{j}}) = 0 \rightarrow 1 = \sum_{i,E_{j}}p_{i,E_{j}}\delta_{E_{i}E_{j}} \\
	& = \sum_{E_{j}}p_{E_{i},i} \rightarrow p_{E_{i},i} = \frac{1}{\Omega(E_{i})},p_{E_{j\neq i},i} = 0.
	\end{aligned}
	\end{equation}  
	We have thus shown, by assuming that $p_{i}$ is a density, that the constraint implies the Dirac delta without invoking the principles of a Maximum Entropy method. This can be readily seen by using Lagrange multipliers. For a function $f(x) = 0$ with constraints $g_{k}(x) = 0$ we solve the equations,
	\begin{equation}
	\label{scalinglagrange}
		\begin{aligned}
	    & \frac{\partial f(x)}{\partial x_{i}} - \sum_{k}\lambda_{k}\frac{\partial g_{k}(x)}{\partial x_{i}} = 0 \\
	    &  \lambda_{l}\bigg(\frac{1}{\lambda_{l}} \frac{\partial f(x)}{\partial x_{i}} -  \frac{\partial g_{l}(x)}{\partial x_{i}} - \sum_{k\neq l}\frac{\lambda_{k}}{\lambda_{l}}\frac{\partial g_{k}(x)}{\partial x_{i}}\bigg) = 0. \\
	    &  \mbox{As }\xrightarrow[\text{lim}]{\lambda_{l}\rightarrow\infty} \mbox{ so } \frac{\partial g_{l}(x)}{\partial x_{i}} = 0.
	    \end{aligned}
	\end{equation}
	Hence in the limit of an infinite Lagrange multiplier, the functional form of the entropy is irrelevant. We are simply solving the constraint equation up to a constant addition, as per \cite{lee,julianandpresse}.
	
	\subsection{Application to the Canonical Ensemble}
	In \cite{julianandpresse}, in order to link the inverse temperature and mean energy constraint, a two step maximisation process is employed, using the decomposition of probabilities and the approximation (due to a large bath size):
	\begin{equation}
	\label{approximationmicrostates}
	\log \Omega(E_{b}-E_{i}) \approx \log \Omega(E_{b}) - \beta E_{i}.
	\end{equation}
	This assumption, which is a consequence of the factorisation of micro states (and thus independence), fully constrains the probability density to be of the exponential form.
	\begin{equation}
	\label{standard}
	p_{i} = \frac{\Omega(E_{b}-E_{i})}{\Omega(E_{b})}
	\end{equation}
	Equating \eqref{approximationmicrostates} and \eqref{standard} we obtain,
	\begin{equation}
    \log p_{i} = \log(\Omega(E_{b}-E_{i})) - \log(\Omega(E_{b})) = -\beta E_{i}.
	\end{equation}
	Generically when considering a large heat bath in contact with an open system, the additive microstates are expanded to first order:
	\begin{equation}
	\begin{aligned}
	   & \Omega(E-E_{\nu})=\Omega(E)\Omega(1-\frac{E_{\nu}}{E}) \\
	   & \propto \Omega(1)+\frac{d\Omega}{dE}\bigg(\frac{-E_{\nu}}{E}\bigg)+\bigg[ \frac{d^{2}\Omega}{dE^{2}}\bigg(\frac{-E_{\nu}}{E}\bigg)^{2}...\bigg].
	    \end{aligned}
	\end{equation}
	Arguments for this truncation \cite{chandler}  often appeal to the independence of random variables for which
	\begin{equation}
	\frac{\sqrt{(E-\langle E \rangle)^{2}}}{E} = \frac{\sqrt{n \mbox{Var}(\epsilon)}}{n\epsilon} \simeq O \left ( \frac{1}{\sqrt{n}} \right ) \rightarrow \underset{n\rightarrow\infty}{0}.
	\end{equation}
	It is worth noting that it can be readily shown \cite{landau_statphys} that independence \emph{alone} is enough to constrain the microstate expansion to have the exact form \eqref{approximationmicrostates}. By Liouville's theorem the density (which is time invariant), must be an equation of the integrals of motion (energy and the three components of (angular) momentum). By independence of sub-systems it must be an additive combination of the integrals of motion. By considering the rest frame of a non-rotating system we recover \eqref{approximationmicrostates} as:
	\begin{equation}
	\label{landau}
	\log p_{i} = \alpha_{i} +\beta E_{i} + \gamma.P_{i} + \delta.\mu_{i} \rightarrow \alpha_{i} +\beta E_{i}.
	\end{equation}
	Where $\alpha,\beta,\gamma,\delta$ are constants and $E_{i}, P_{i}, \mu_{i}$ are the energy and components of momentum and angular momentum respectively. Equating equations \eqref{standard} and \eqref{landau} reproduces equation \eqref{approximationmicrostates}. We note that a corresponding derivation holds for quantum statistics. The quantum analogue of Liouville's theorem is that the statistical matrix (QM equivalent of the distribution function) must commute with the Hamiltionian, in the energy representation it is thus diagonal. Applying this to quasi-closed subsystems one has \cite{landau_statphys}
	\begin{equation}
	\log w_{n}^{i}=\alpha^{i}+\beta E_{n}^{i}
	\end{equation}
	for each subsystem $i$. This is equivalent to Equation \eqref{landau}.
	
	\subsection{Contributions of this paper}	
	We look to fill a gap in the literature by analytically deriving the mean value constraints (which by the application of MaxEnt lead to the canonical ensembles \cite{Jaynes1967}). We do this via two equivalent methods. We initially proceed by demonstrating that independence follows naturally from the MaxEnt framework and then applying independence to an arbitrary probability distribution over the macro-variables of interest (energy, particle number). By the central limit theorem in the thermodynamic limit ($N \rightarrow \infty)$, we recover the mean value constraints. We also derive the same mean value constraints from the underlying quantum field theory. By positing a time-invariant underlying Lagrangian and using the quantum version of Noether's theorem (namely the Ward Identities), one recovers the mean energy constraint (leading to the canonical ensemble by MaxEnt). By positing a free field scalar Lagrangian (no interactions and hence independent) by Ehrenfest's theorem we then recover a particle number constraint (leading to the grand canonical ensemble by MaxEnt). We complete the gap we perceive in the literature, allowing the derivation of the ensembles without the need for additional assumptions such as ergodicity, metric-transitivity and equal a priori probabilities. Further, noting that quantum mechanics has been shown to be the Bayesian theory extended to complex Hermitian spaces \cite{qmhermitian} and that nanoscale systems are routinely investigated \cite{singlemoleculedna}, we regard it as essential to have a fully Bayesian formulation of statistical mechanics, free from infinite mental assembly arguments. By creating an equivalence between free field Lagrangians and statistical independence, we allow the analysis of dependent and small-scale systems and make a tentative link towards the work of Tsallis and its success in optical lattices \cite{latticetsallis,optiontsallis}, as well as opening up possible application domains such as finance \cite{qreview}.

	\section{Independence and Entropy}
	\subsection{Independence}
	It is often assumed, due to weak coupling, that the energy (of a statistical mechanical system) is additive and the number of states can thus be factorised. This necessarily implies statistical independence. It can be shown that due to the decomposition of probabilities \cite{cox} that the joint entropy is maximal for independent variables; 
    \begin{equation}
    \label{mutualinfo}
     H(X,Y)  = H(X) + H(Y)
    - I(X;Y)  \leq H(X) + H(Y),
      \end{equation}
which follows from the fact that the mutual information, $I(X;Y)$, in equation \eqref{mutualinfo}, as a KL-divergence (relative entropy), is never negative:
	\begin{equation}
	\begin{aligned}
	 &	I(X;Y) = -\sum_{x,y}p(x,y)\log\frac{p(x)p(y)}{p(x,y)} \\
	 & = -\mathbb{E}\bigg[\log\frac{p(x)p(y)}{p(x,y)}\bigg] \geq -\log \bigg(\mathbb{E}\bigg[\frac{p(x)p(y)}{p(x,y)}\bigg]\bigg) = 0,
	 	\end{aligned}
	\end{equation}
	in which $\mathbb{E}(.)$ is the expectation operator.
	Hence we show that the assumption of independent random variables is consistent with the principle of maximum entropy.

	\subsection{Application to Statistical Mechanics}
	\label{independentderivation}
We now consider the application of independence, maximum entropy and the central limit theorem (appendix \ref{CLT}) to an ensemble of particles. For a system of $N$ particles, the combinatorial most likely (maximum entropy) state is for them to be independent (this follows from Equation \eqref{mutualinfo}). 

Let us assume that there exists some (probability) distribution $p(\vec{x_{1}}...\vec{x_{n}})$, (where $\vec{x_{i}}$ is the state vector of each individual particle), that describes the complete microstate of the whole ensemble of particles. The entropy is hence,
\begin{equation}
\begin{aligned}
\label{systementropy}
   & S= -\int p(\vec{x_{1}}...\vec{x_{n}})\ln p(\vec{x_{1}}...\vec{x_{n}})d(\vec{x_{1}}...\vec{x_{n}})\\
   & -\alpha\bigg[\int p(\vec{x_{1}}...\vec{x_{n}})d(\vec{x_{1}}...\vec{x_{n}}) -1\bigg] \\
   & - \sum_{i,m}\lambda_{i,m}\bigg[\int p(\vec{x_{1}}...\vec{x_{n}}) \epsilon(\vec{x_{1}}...\vec{x_{n}})^{m}d(\vec{x_{1}}...\vec{x_{n}}) - \mu_{i,m}\bigg].
    \end{aligned}
\end{equation}
Where $S$ in the above is the system entropy, $\alpha$ and $\lambda$ are the Lagrange multipliers and $\epsilon(\vec{x_{i}})$ is the energy of the system, which depends on all the particles and we characterise by its moments, which are defined by the power $m$ in the above. By being maximally entropic, we consider our particles to be independent and hence the probabilities factorise, thus $p(\vec{x_{1}}...\vec{x_{n}})=\Pi_{i}p(\vec{x_{i}})$. We can therefore write $\epsilon(\vec{x_{1}}...\vec{x_{n}}) = \sum_{i}\epsilon(\vec{x_{i}})$ and note that all cross correlations are 0, as $\langle \epsilon(\vec{x_{i}})\epsilon(\vec{x_{j}}) \rangle_{i\not = j} = 0$.
Further, as there is no way to distinguish one particle from another, they are also identical probability distributions, 
\begin{equation}
\begin{aligned}
    & S = \sum_{i}\bigg(-\int p(\vec{x_{i}})\ln p(\vec{x_{i}})d\vec{x_{i}}\\
    & - \alpha\bigg[\int p(\vec{x_{i}})d\vec{x_{i}} - 1 \bigg]\\
    & -\sum_{m>1}^{\infty}\lambda_{m}\bigg[\int p(\vec{x_{i}})\epsilon(\vec{x_{i}})^m d\vec{x_{i}} - \mu_{i} \bigg] \bigg).
    \end{aligned}
\end{equation}
Writing the MaxEnt constraints in terms of cumulants and noting that $ n = \sum_{i}$ and $S_{N} = \sum_{i}\epsilon_{i}/n$, we obtain:
\begin{equation}
 \sum_{i}C_{m}(\vec{\epsilon_{i}})=nC_{m}(\sum_{i}\vec{\epsilon_{i}}/n) = \sum_{i}C_{m}(S_{N}),
\end{equation}
Hence we can write $S$ as
\begin{equation}
\begin{aligned}
S = \sum_{i}\bigg(-\int p(\vec{x_{i}})\ln p(\vec{x_{i}})d\vec{x_{i}}
-\alpha\bigg[\int p(\vec{x_{i}})\vec{x_{i}})d\vec{x_{i}} - 1 \bigg]\\ -\sum_{m>1}^{\infty}\lambda_{m}\bigg[C_{m}(S_{N})-\mu_{i}\bigg] \bigg)
\end{aligned}
\end{equation}
We now invoke the central limit theorem (appendix \ref{CLT}). It can be shown that for $N$ identical independent variables the cumulant of their average varies as $C_{m}(S_{N}) =C_{m}[\epsilon(\vec{x_{i}})]N^{1-m}$ and hence for each particle we may solve an effective mean field given as:
\begin{equation}
\label{meanfield}
\begin{aligned}
S = -\int p(\vec{x})\ln p(\vec{x})d\vec{x_{i}} - \alpha\bigg[\int p(\vec{x}))d\vec{x_{i}} - 1 \bigg]\\ -\sum_{m>1}^{\infty}\lambda_{m}N^{1-m}\bigg[C_{m}[\epsilon(\vec{x_{i}})]-\mu'_{i}\bigg]
\end{aligned}
\end{equation}
So long as the cumulants (which can easily be written in terms of moments, see appendix \ref{dirac}) are not infinite, our only non-zero constraint (as $\mu'_{i}$ is a constant) in the $N\rightarrow\infty$ limit is for $m = 1$. We note that the $m=0$ term is our normalisation constraint, treated separately with its Lagrange multiplier $\alpha$. We hence recover a system with  a mean value constraint:
\begin{equation}
\begin{aligned}
S = -\int p(\vec{x})\ln p(\vec{x})d\vec{x_{i}} - \alpha\bigg[\int p(\vec{x})d\vec{x_{i}} - 1 \bigg]\\ -\lambda_{1}\bigg[C_{1}[\epsilon(\vec{x_{i}})]-\mu'_{i}\bigg],
\end{aligned}
\end{equation}
in which $C_{1}[\epsilon(\vec{x})] = \langle \epsilon(\vec{x}) \rangle = \int p(\vec{x})\epsilon(\vec{x})\thinspace dx$. We note that through this formulation we have derived the mean energy value constraint via assuming independence (which is maximally entropic) and the central limit theorem (\ref{CLT}) in the thermodynamic limit ($N \rightarrow \infty$). Whilst we have applied the method to energy, the formalism naturally extends to other variables of interest, such as particle number. In such extended configurations we have an extra set of Lagrange multipliers in equation \eqref{systementropy} over the set of moment expectations $\langle N(\vec{x_{1}..x_{n}})^{m}\rangle$. The principle of independence can be readily applied again, in this case to independence between energy and particle number. From this, e.g., we recover a mean particle number constraint along with a mean energy constraint.

The importance of the approach lies in the elegant equivalence between the ensemble and maximum entropy derivations of statistical mechanics under the independence assumption (present in the factorisation of microstates and the additivity of energy). 

We further note that convergence can be extended to $m$-dependent variables \cite{stein1972}, where we have a bounded error to the central limit theorem of $\mathcal{O}(N^{-1/2})$ under the condition of a finite 8th moment ($E(X^8)<\infty$) and a finite variance of the sum of the variables. For exponentially decaying variables \cite{convergenceclt} with a finite 4th moment ($E(X^4)<\infty$), strong mixing and regularity leads to an error to the central limit theorem of $\mathcal{O}(N^{-1/2}\log(N^{1/2}))$. We note that the framework developed in the preceding section is robust to weak dependence and remains exact in the thermodynamic limit ($N \rightarrow \infty$), which formalises the meaning behind sub-systems being ``approximately independent" \cite{landau_statphys}.

We also suggest that this formalism is indicative of how to handle small scale systems (where the thermodynamic limit is not appropriate) and highly dependent interacting systems (where the independence assumption no longer holds). Note that in Equation \eqref{meanfield},  larger higher-order moment information (coming via the cumulant) is equivalent to smaller moment information for a smaller particle number $N$. This suggests an equivalence between large dependent systems and small independent systems. We leave the rigorous formulation and applications thereof for future work, but note that in many systems of interest the number of observations or agents is far smaller than those of a typical thermodynamic system (namely of order $10^{27}$). We further note that for small systems, the total probability mass in the immediate vicinity of the maximum entropy prediction shrinks ($p \propto \exp(NS)$) and so the problem of averaging over states of lower entropy becomes important. In the discussion we demonstrate how both Tsallis q-exponentials and q-gaussians emerge naturally from classical statistical mechanics in the limits of small heat baths and rest masses. 
	
\section{Microcanonical Ensemble}
	
	For a time invariant Lagrangian, by Noether's theorem (see Appendix \ref{noether}) we have a fixed energy. This necessarily implies a delta distribution around the system energy value $E_{i}$. Constraining the distribution so that its sum (or integral) is unity leads to:
	\begin{equation}
	\begin{aligned}
	& \delta \bigg(-\int_{E=E_{i}} p(x)\ln p(x)dx - \alpha\bigg[\int_{E=E_{i}} p(x)dx - 1\bigg]\bigg) = 0 \\
	& \rightarrow p(x) = \exp^{-[1+\alpha]} = \frac{1}{\Omega(E_{i})} = \frac{\delta(E-E_{i})}{\Omega(E)}.
	\end{aligned}
	\end{equation}
    It can be thence shown that for the infinite set of constraints which a distribution of fixed energy must satisfy (all central moments being 0), that the unique distribution to satisfy these constraints is the Dirac delta (see Appendix \ref{dirac}). This derivation does not postulate ergodicity, metric transitivity and equal a priori probabilities.
	
	\section{The Canonical Ensemble as a Maximum Entropy  Distribution}
	\label{maxentcanonical}
	
	For a system with a conserved mean energy, we maximize the entropy functional with an additional Lagrange multiplier,
	\begin{equation}
	\label{canonical}
	\begin{aligned}
	\delta S = \frac{\partial}{\partial p_{j}} \bigg[ -\sum_{i} p_{i}ln p_{i} - \alpha\bigg(\sum_{i}p_{i}-1\bigg) \\ -\beta\bigg(\sum_{i}p_{i}\epsilon_{i}-\langle \epsilon \rangle\bigg)\bigg]=0
	\end{aligned}
    \end{equation}
	This can be solved  to give the canonical ensemble as below,
	\begin{equation}
	\label{canonicalensemble}
	p_{i} = \frac{e^{-\beta\epsilon_{i}}}{\sum_{i}e^{-\beta\epsilon_{i}}} = \frac{e^{-\beta\epsilon_{i}}}{Z(\beta)}.
	\end{equation}
	Where the Lagrange multiplier $\beta$ is $\langle \epsilon \rangle^{-1}$ from the mean value constraint (the third expression in Equation \eqref{canonical}).
	
\subsection{Mean energy constraint}
In  Section \ref{independentderivation} we derived the mean energy constraint explicitly using arguments of independence and the central limit theorem in the thermodynamic limit. We now show that equivalently, the constraint can be derived from the underlying quantum field theory. Generalising Noether's theorem from classical field theory to give a quantum field theoretic version, we derive the Ward identities (Appendix \ref{ward}). The end result is that we no longer have a conserved current but a conserved \emph{expected} current,
	\begin{equation}
    \partial^{\mu}\langle j_{\mu}(x) \rangle = 0.
	\end{equation}
In the case of continuous time translation symmetry we would now also have a conserved energy expectation. Under the MaxEnt framework this also leads to eqn \eqref{canonicalensemble}.
	
\section{Grand Canonical Ensembles from Maximum Entropy}	
		
Generalising Equation \eqref{canonical} to a two-dimensional probability distribution over both energy and particle number, with mean constraints on both energy and particle number we have,
	\begin{equation}
	\begin{aligned}
	& \frac{\partial}{\partial p_{k,l}}\bigg[-\sum_{i,j}p_{i,j}\ln(p_{i,j})-\alpha\bigg(\sum_{i,j}p_{i,j}-1\bigg)\\
	&-\beta\bigg(\sum_{i,j} p_{ij}\epsilon_{i} - \langle \epsilon \rangle\bigg) -\gamma\bigg(\sum_{i,j} p_{i,j}n_{j} - \langle n \rangle\bigg)\bigg] = 0.
	\end{aligned}
	\end{equation}
therefore,
\begin{equation}
p_{ij} = \frac{e^{-\beta(\epsilon_{i}+\frac{\gamma}{\beta}n_{j})}}{\sum_{i,j}e^{-\beta(\epsilon_{i}+\frac{\gamma}{\beta}n_{j})}}.
\end{equation}
upon making the substitution $\gamma/\beta = -\mu$ we arrive at the celebrated grand canonical ensemble probability distribution,
\begin{equation}
	p_{ij} = \frac{e^{-\beta(\epsilon_{i}-\mu n_{j})}}{Z(\beta)}.
\end{equation}
	
\subsection{Conserved mean particle number}
In this section, we derive the constraint for a fixed particle number using Ehrenfest's theorem and a free field theory (which posits underlying independence between the particles). We retain the constant energy expectation from the previous section.

\subsection{Ehrenfest's Theorem}
We start by quoting Ehrenfest's theorem,
	\begin{equation}
	\begin{aligned}
	\frac{d}{dt}\langle \hat{O} \rangle = \frac{d}{dt}\int \psi^{*}\hat{O}\psi \\
	= \int \psi \frac{\partial \hat{O}}{\partial t}\psi + \int \frac{\partial \psi^{*}}{\partial t}\hat{O}\psi + \int \psi^{*}\hat{O}\frac{\partial \psi^{}}{\partial t}.
	\end{aligned}
	\end{equation}
Using Schroedinger's equation,
	$H|\psi\rangle = i\hbar\frac{\partial |\psi\rangle}{\partial t}$,
we obtain,
	\begin{equation}
	\frac{d}{dt}\langle \hat{O} \rangle = \left < \frac{\partial \hat{O}}{\partial t} \right > + \frac{1}{i\hbar}\langle\psi|[\hat{O},\hat{H}]|\psi\rangle.
	\end{equation}

\subsection{Scalar Klein-Gordon}
The simplest relativistic-free theory is that defined by the classical Klein-Gordon (KG) equation, in which the Lagrangian, $\mathcal{L}$, is given as,
	\begin{equation}
	\mathcal{L} = \frac{1}{2}(\partial_{\mu}\phi)^{2} -\frac{1}{2}m^{2}\phi^{2}.
	\end{equation}
It can be shown by writing out the fields, number operator and conjugate momenta in terms of the creation and annihilation operators (see Appendix \ref{scalargordon}) that for a free field theory, the Hamiltonian commutes with the number operator, i.e. $[N,H]=0$. Applying Ehrenfest's theorem to the number operator we see that,
	\begin{equation}
	\frac{d}{dt}\langle \hat{N} \rangle = \left < \frac{\partial \hat{N}}{\partial t} \right > + \frac{1}{i\hbar}\langle\psi|[\hat{N},\hat{H}]|\psi\rangle = 0,
	\end{equation}
as the number operator does not explicitly depend on time and its commutator is 0. Hence,
	\begin{equation}
	\langle N \rangle = k.
	\end{equation}
This completes our derivation. We see that for a time-invariant, non-interacting Lagrangian (one in which the number operator commutes with the Hamiltonian), in the large number limit (this causes the maximum entropy solution to dominate the probability mass), the exponentially most likely solution is the grand canonical ensemble. 
	
\section{Extension to Interacting Complex Lagrangians}
	\label{interactinglagrangian}
A Lagrangian of the form $\mathcal{L} = \partial_{\mu}\psi^{*}\partial^{\mu}\psi - V(|\psi|^{2})$ is invariant under the change $\psi \rightarrow e^{i\alpha}\psi$ or $\delta \psi = i\alpha\psi$. $\delta \psi^{*} = i\alpha\psi^{*}$. Due to the Ward identities (see Appendix \ref{ward}), expanding the complex field operators and the classical field momentum as a sum of plane waves (see Appendix \ref{U(1)}) we have a conserved mean quantum charge $\langle Q \rangle = \langle N_{c}-N_{b}\rangle$ in the interacting theory. This means, in the thermodynamic limit, following a similar argument to that of Section \ref{maxentcanonical} we would expect an exponential over the particle number of the form,
\begin{equation}
	P_{b,c} = \frac{1}{Z}\exp[(N_{c}-N_{b})/\langle N_{c}-N_{b} \rangle].
\end{equation}
\\
And hence we see that even for strongly interacting systems with symmetries we expect to be able to characterise these systems by exponential distributions over the conserved expectations induced by the symmetries. 

\section{Conclusion and Discussion}
In this paper we have derived all the classical statistical mechanical ensembles without the need for ergodicity, metric transitivity or the postulate of equal a priori probabilities. By further exploiting the independence assumption (already present in the ensemble theory approach) and applying the central limit theorem in the thermodynamic limit, we derived the mean value constraints. By the method of maximum entropy these constraints reproduce the canonical and grand-canonical ensembles. The mean value constraints were also derived by the application of the Ward identities and Ehrenfest's theorem to a time invariant free-field Lagrangian. That we arrive at the same prediction under two completely different formalisms builds on the work of Jaynes \cite{Jaynes1967} and Benavoli \cite{qmhermitian}, displaying the extent to which Physics can be considered as a theory of inference. 	
Open questions still remain, such as how to extend the formalism to predict probability distributions of systems characterised by non-time invariant interacting Lagrangians. Our result in Section \ref{interactinglagrangian}, exploiting the U(1) symmetry for an interacting Lagrangian, perhaps offers a starting point. Analogies between higher-order Lagrangians and statistical dependence have also not been explored in this paper.
		
By considering statistical results for weakly dependent variables \cite{stein1972,convergenceclt}, we have shown the robustness of the conclusion under certain types of dependence in the thermodynamic limit. Unanswered questions include how to cope in situations of extreme dependence and in the small number limit. We further note that the informativeness of the MaxEnt solution is dependent on the sharpness of the predictive distribution around its most entropic value; for small systems the contribution of lower entropy solutions will be significant and should be taken into account.

\textit{The value of Tsallis entropies: }
The BSG entropic functional (Equation \eqref{BSG}) has been extended to a q-entropic functional \cite{tsallis_2009}. It has been used to better model financial returns and price options \cite{optiontsallis} as well as predict the distribution in optical lattices \cite{latticetsallis}. Its free paramater, $q$, defines the fluctuations of the generalised temperature (following a gamma distribution) \cite{whatistsallisq} and can be derived from Langevin equations with fluctuating forces \cite{beck_2001}. It has been criticised for failing to reproduce basic thermodynamical laws such as Kirchoff's and Boltzmann's law for anything other than $q=1$ (the value which reduces it to the BSG entropy) \cite{tsalliscritiquethermal} and for introducing unwanted bias not warranted by the data \cite{tsalliscritiquedata}. We simply show that, in a much more restrictive sense ($q \leq 1$), the Tsallis q-exponential and q-Gaussian can be derived from generic statistical mechanics in the limit of small particle number and negligible rest mass.
	
We start by writing the total number of microstates of an ideal gas as,
    \begin{equation}
	\Omega(E,N,V) = \frac{1}{\hbar^{3N}}\int_{V}\Pi_{i}d\vec{x_{i}}\int_{E}\Pi_{i}d\vec{p_{i}} = \frac{V^{N}}{\hbar^{3N}}\int_{E}\Pi_{i}d\vec{p_{i}}.
	\end{equation}
The momenta are constrained by the usual relation $(1/2m)\sum_{i}(\vec{p_{i}}^{2})=E$, so we have a $3N$ dimensional hypersphere of radius $\sqrt{2mE}$ and accessible momentum space is $RdR$. Combining this with the equation of volume and surface area of an $n$-dimensional hypersphere,
	\begin{equation}
	V_{n}(R) = \frac{\pi^{n/2}}{(n/2)!}R^{n}, \thinspace S_{n}(R) = \frac{2\pi^{n/2}}{\gamma(n/2)!}R^{n-1}, 
	\end{equation}
we have
	\begin{equation}
	\Omega(E,N,V) = \frac{V^{N}}{\hbar^{3N}}\frac{(2\pi mE)^{3N/2}}{(3N/2-1)!}\frac{\delta E}{E}.
	\end{equation}
Hence, as discussed in \cite{caticha2012entropic},
	\begin{equation}
	\label{microstatestotal}
	p_{i} = \frac{\Omega(E_{b}-E_{i})}{E_{b}} \propto (1-E_{i}/E_{b})^{3N/2-1},
	\end{equation}
	which is equivalent to the Tsallis q-exponential 
	\begin{equation}
	    p_{i} \propto (1-\lambda E_{i})^{1/1-q}.
	\end{equation}
We note that the constraint $\sum_{i}p_{i}^{2} = 2mE$ is a special case of the special relativistic invariant equation $E^{2} = p^{2}c^{2} + M_{0}^{2}c^{4}$, where $E$ is the energy of the particle, $p$ is the momentum, $M_{0}$ the particle rest mass and $c$ is the speed of light. Adopting natural units where $c=1$ we have $\sum_{i}p_{i}^{2} = E^{2}-M_{0}^{2}$. Analogously to \eqref{microstatestotal} we generate a distribution which is q-exponential in the Newtonian limit and q-Gaussian\footnote{Equation \ref{notquiteqgaussian} is not the q-Gaussian as $E\not < 0$. It is a truncated half q-Gaussian.} in the ultra relativistic limit. 
	\begin{equation}
	\label{notquiteqgaussian}
	p_{i} \propto (1-\lambda(E_{i}^{2}-M_{0,i}^{2}))^{1/1-q}.
	\end{equation}
The notion of mass emerging generally as a constraint on mean path square displacement (imposing smoothness) \cite{Gonzalez2014} gives insight as to which distribution may be appropriate in generic inference.  Whilst it is promising that we have made a link between q distributions and the departure from the thermodynamic limit ($N \not \to \infty$) and the Newtonian limit ($(E^{2}-M_{0}^{2}c^{4})/c^2 \not \approx 2mE$), we note that the for the above formalism 
	\begin{equation}
		q = 1 - (DN/2-1)^{-1} \leq 1,
	\end{equation}
where $D$ is the dimension of the system. Hence none of the q-distributions describing financial, optical lattice or turbulent phenomena \cite{optiontsallis,latticetsallis,beck_2001} can be described by this. However, it derives the probability distribution of a bounded variable, used in Biology, Engineering and Physics \cite{boundedqleq1} as a function of bath size.
	
	\appendix
		\section{Ergodicity and Metric Transitivity}
		\label{ergodicity}
	Generically, the functional form of the micro-canonical ensemble is derived under the assumption that for an isolated (constant energy) system in equilibrium, all accessible microstates are equally likely. This restricts us to the probability distribution \cite{chandler} given by,
	\begin{equation}
	    P_{s} = \frac{\delta(E_{s}-U)}{\Omega(E)} = \frac{1}{\Omega(U)}.
	\end{equation}
	This starting point can be derived by assuming ergodicity and metric transitivity. Metric transitivity ensures that the trajectories can move freely on the energy surface (all states of equal energy are accessible), whilst ergodicity ensures that the time the trajectory spends in a region phase space is proportional to its volume (we spend an equal amount of time in all the states of identical energy). As we have nothing to distinguish one region from another, so the best choice we can make is to assume that the probability $P(R_{e})$ of finding the system in $R_{E}$ is equal to the fraction of the energy surface occupied by $R_{E}$ (this is known as the equal a priori postulate of statistical mechanics). Thus,
	\begin{equation}
	    P(R_{E}) = \frac{1}{\sum(E)}\int_{R_{E}}dS_{E}=\frac{\sum(R_{E})}{\sum(E)},
	\end{equation}
	which when correctly normalised gives (on the accessible energy surface),
	\begin{equation}
	    \rho(X^{N},S_{E}) = \frac{1}{\sum(E)}.
	\end{equation}
		\section{Noether's Theorem}
	\label{noether}
	For any continuous symmetry we can work infinitesimally using $\delta \phi_{a}(x) = X_{a}(\phi)$. To keep the action invariant we must change the Lagrangian by a total derivative. This is because the equations of motion are derived from extremising the action and hence for fixed boundaries any added constant value $F(b)-F(a)$ is irrelevant. Hence $\partial \mathcal{L} = \partial_{\mu}F^{\mu}$, for some set of functions $F^{\mu}(\phi)$. For an arbitrary transformation of the fields $\partial \phi_{a}$,
	\begin{equation}
	\begin{aligned}
	& \partial \mathcal{L} = \frac{\partial \mathcal{L}}{\partial \phi_{a}}\delta \phi_{a} + \frac{\partial \mathcal{L}}{\partial(\partial_{\mu}\phi_{a})}\partial_{\mu}(\delta \phi_{a})\\
	& \partial \mathcal{L} = \bigg[\frac{\partial \mathcal{L}}{\partial \phi_{a}} - \partial_{\mu}\frac{\partial \mathcal{L}}{\partial(\partial_{\mu}\phi_{a})}\bigg]\partial\phi_{a}+\partial_{\mu}\bigg(\frac{\partial \mathcal{L}}{\partial(\partial_{\mu}\phi_{a})}\partial\phi_{a}\bigg).
	\end{aligned}
	\end{equation}
	The first term in the above is 0 when the equations of motion are satisfied, so by equating this with our total derivative we have,
	\begin{equation}
	\partial_{\mu}j^{\mu} = \partial_{\mu}\bigg[ \frac{\partial \mathcal{L}}{\partial(\partial_{\mu}\phi_{a})}X_{a}(\phi) - F^{\mu}(\phi)\bigg] = 0.
	\end{equation}
	
	\subsection{Energy-Momentum Tensor}
	
	If we consider the infinitesimal translation, $x^{v} \rightarrow x^{v} - \epsilon^{v}$, $\phi_{a}(x) \rightarrow \phi_{a}(x) + \epsilon^{v}\partial_{v}\phi_{a}(x)$, then we have a variation in the Lagrangian of $\mathcal{L}(x) \rightarrow  \mathcal{L}(x) + \epsilon^{v}\partial_{v}\mathcal{L}(x) = \mathcal{L}(x) + \epsilon^{v}\partial_{\mu}(\delta^{\mu}_{v}\mathcal{L}(x))$. Using Noether's theorem we get four conserved currents 
	\begin{equation}
	j^{\mu}_{v} = \frac{\partial \mathcal{L}}{\partial(\partial_{\mu}\phi_{a})}\partial_{v}\phi_{a} - \delta^{\mu}_{v}\mathcal{L} \equiv T^{\mu}_{v}.
	\end{equation}
	For a Lagrangian of the form for a real scalar field, $\frac{1}{2}\eta^{\mu v}\partial_{\mu}\phi \partial_{v} \phi - \frac{1}{2}m^{2}\phi^{2}$, we have an energy momentum tensor
	\begin{equation}
	T^{\mu v} = \partial^{\mu}\phi \partial^{v}\phi - \eta^{\mu v}\mathcal{L}.
	\end{equation}
	The 0th component of the current is conserved, giving for conserved momenta 
	\begin{equation}
	P^{i} = \int d^{3}x  T^{0i}.
	\end{equation}
	For $i = 0$ (the index of time) we have energy and for $i = 1,2,3$ we have the momenta in the $x,y,z$ directions. 
	
	\section{Mean energy from Quantum Mechanics}
	\label{ward}
	\subsection{From Classical to Quantum}
	We now generalise Noether's seminal result from classical field theory to give a quantum field theoretic version of the result. Completing the same procedure as before, in the notation of Riemann geometry we have:
	\begin{equation}
	\label{actionvariation}
	\begin{aligned}
	& \delta_{\epsilon}S[\phi] = 0 = - \int_{M}*j \wedge d\epsilon = \int_{M}g^{\mu v}j_{\mu}(x)\partial_{v}\epsilon(x)\sqrt{g}d^{d}x\\
	& \rightarrow d*j = \partial_{v}(\sqrt{g}g^{\mu v}j_{v}) = 0,
	\end{aligned}
	\end{equation}
	where  $*$ represents the Hodge dual,  $\wedge$ is the wedge product and  $d\epsilon$ is the exterior derivative.
	
	\subsection{Ward's Identities}
	Supposing that a local transformation $\phi \rightarrow \phi'(\phi)$ leaves the product of the action and the path integral measure invariant, so:
	\begin{equation}
	\label{wardeq}
	\mathcal{D}\phi e^{-S^{eff}_{\Lambda}[\phi]} = \mathcal{D}\phi' e^{-S^{eff}_{\Lambda}[\phi']}.
	\end{equation}
	On a compact manifold $M$, by relabelling $\phi$ by $\phi'$ and using the identity \eqref{wardeq}, we have that the correlation functions obey the relations:
	\begin{equation}
	\langle \mathcal{O}_{1}(\phi(x_{1}))...\mathcal{O}_{n}(\phi(x_{n}))\rangle = \langle \mathcal{O}_{1}(\phi'(x_{1}))...\mathcal{O}_{n}(\phi'(x_{n}))\rangle.
	\end{equation}
	Hence symmetries impose selection rules on the operators we can insert, if we wish to obtain a non-zero correlator. Computing the path integral in terms of the primed quantities where $\phi' = \phi + \delta_{\epsilon}\phi$ yields:
	\begin{equation}
	\label{partitionqft}
	\begin{aligned}
&	\mathcal{Z} = \int \mathcal{D}\phi'e^{-S^{eff}_{\Lambda}[\phi']} = \int \mathcal{D}[\phi+\delta_{\epsilon}\phi]e^{-S_{\Lambda}^{eff}[\phi+\delta_{\epsilon}\phi]}\\
& = \int \mathcal{D}\phi[1+\delta_{\epsilon}S]e^{-S[\phi]}(1-\delta_{\epsilon}\phi\delta_{\epsilon}S...).
\end{aligned}
	\end{equation}
	The last term in \eqref{partitionqft} is second order and is thus neglected. Using Equation \eqref{actionvariation} this reduces to,
	\begin{equation}
	\begin{aligned}
	& \int \mathcal{D}\phi'e^{-S^{eff}_{\Lambda}[\phi']} = \int \mathcal{D}\phi e^{-S^{eff}_{\Lambda}[\phi]}\bigg[1-\int_{M}*j\wedge d\epsilon\bigg]\\
	& = \int \mathcal{D}\phi e^{-S^{eff}_{\Lambda}[\phi]}.
	\end{aligned}
	\end{equation}
	Hence, as the partition function (cast here in path integral form) calculates sample expectations, we obtain:
	\begin{equation}
	\label{wardmeanconstraint}
	0 = -\int_{M}*\langle j(x) \rangle \wedge d\epsilon = \int_{M}d*\langle j(x)) \rangle \rightarrow \partial^{\mu}\langle j_{\mu}(x) \rangle = 0.
	\end{equation}
	In the case of  continuous time translation symmetry we would now have a conserved energy expectation. It is worth noting that if Equation \eqref{wardmeanconstraint} applied to the conserved charge of Noether's theorem (see Appendix \ref{noether}) we obtain a mean conserved charge.
	
	\subsection{Scalar Klein Gordon system}
	\label{scalargordon}
	The simplest relativistic free theory leads to the classical Klein-Gordon (KG) equation for the Langragian:
	\begin{equation}
	\mathcal{L} = \frac{1}{2}(\partial_{\mu}\phi)^{2} -\frac{1}{2}m^{2}\phi^{2},
	\end{equation}
	\begin{equation}
	\label{kgreal}
	\thinspace \mathcal{H} = p\dot{q}-\mathcal{L} = (\partial_{0}\phi)^{2}-\mathcal{L} = \frac{1}{2}\pi^{2}+\frac{1}{2}(\nabla \phi)^{2}+\frac{1}{2}m^{2}\phi^{2}.
	\end{equation}
	Writing out the fields and their conjugate momenta in terms of the creation and annihilation operators, $w_{p} = \sqrt{|p|^{2}+m^{2}}$ and $p_{\mu} = (w_{p},\textbf{p})$, $[a(p),a(q)]=0$ $\&$ $[a(p), a^{\dag}(q)] = (2\pi)^{3}2w_{p}\delta^{3}(\textbf{p}-\textbf{q})$ we obtain
	\begin{equation}
	\phi(x,t) = \int \frac{d^{3}p}{(2\pi)^{3}}\frac{1}{2w_{p}}(a(p)e^{-ipx}+a^{\dag}(p)e^{ipx}), \nonumber
	\end{equation}
	\begin{equation}
	\pi(x,t) = \dot{\phi}(x,t) = -i \int \frac{d^{3}p}{(2\pi)^{3}}\frac{1}{2}(a(p)e^{-ipx}-a^{\dag}(p)e^{ipx}). \nonumber
	\end{equation}
	By substituting these relations into the Hamiltonian of Equation \eqref{kgreal} and by placing all the annihilation operators on the left (normal ordering) we obtain the Hamiltonian in terms of the creation/annihilation operators,
	\begin{equation}
	:H: \mbox{ } = \int \frac{d^{3}p}{(2\pi)^{3}}\frac{1}{2}a^{\dag}(p)a(p),
	\end{equation}
	which commutes with the number operator $[N,H]=0$. Hence,
	\begin{equation}
	N = \int \frac{d^{3}p}{(2\pi)^{3}}\frac{1}{2w_{p}}a^{\dag}(p)a(p).
	\end{equation}
	This can be seen to count the number of particles in a state by the use of its commutation relation $[N,a^{\dag}(q)] = a^{\dag}(q)$ on the state $N|p_{1},...,p_{n}\rangle$ where $n|0\rangle=0$. 
	
	\section{U(1) symmetry for complex Lagrangians}
	\label{U(1)}
	A Lagrangian of the form $\mathcal{L} = \partial_{\mu}\psi^{*}\partial^{\mu}\psi - V(|\psi|^{2})$ is invariant under the change $\psi \rightarrow e^{i\alpha}\psi$ or $\delta \psi = i\alpha\psi$. $\delta \psi^{*} = i\alpha\psi^{*}$.
	Taking our general equation for the conserved current, we obtain an unchanged Lagrangian, hence $F^{\mu} = 0$. However, the Lagrangian now depends on both $\psi$ and its conjugate $\psi^{*}$ thus we have the expression for $\partial_{\mu}j^{\mu}$ given as,
	\begin{equation}
	\partial_{\mu}\bigg[ \frac{\partial\mathcal{L}}{\partial (\partial_{\mu}\psi)}\delta \psi + \frac{\partial\mathcal{L}}{\partial (\partial_{\mu}\psi^{*})}\delta \psi^{*}\bigg] = \partial_{\mu}\bigg[i(\partial^{\mu}\psi^{*})\psi - i\psi^{*}(\partial^{\mu}\psi)\bigg].
	\end{equation}
	This acts as a conservation of charge or particle number. By expanding the complex field operator and the classical field momentum as a sum of plane waves we obtain,
	\begin{equation}
	\psi = \int \frac{d^{3}p}{(2\pi)^{3}}\frac{1}{\sqrt{2E_{\vec{p}}}}\bigg(b_{\vec{p}}e^{+i\vec{p}.\vec{x}}+c^{\dag}_{\vec{p}}e^{-i\vec{p}.\vec{x}}\bigg), \nonumber
	\end{equation}
	\begin{equation}
	\pi = \int \frac{d^{3}p}{(2\pi)^{3}}i\sqrt{\frac{E_{\vec{p}}}{2}}\bigg(b^{\dag}_{\vec{p}}e^{-i\vec{p}.\vec{x}}-c_{\vec{p}}e^{+i\vec{p}.\vec{x}}\bigg), 
	\end{equation}
	which satisfy the relations $[\psi(\vec{x}),\pi(\vec{y})] = i\delta^{3}(\vec{x}-\vec{y})$ and $[\psi(\vec{x}),\pi^{\dag}(\vec{y})] = 0$. This is thus equivalent to the relations $[b_{\vec{p},\vec{q}},c_{\vec{p},\vec{q}}] = (2\pi)^{3}\delta^{3}(\vec{p}-\vec{q})$, the same for $c$ and all other commutators (or anti-commutators if the underlying statistics are Fermi-Dirac instead of Bose-Einstein) being 0. The classical field momentum $\pi = \partial \mathcal{L}/\partial \dot{\psi} = \psi^{*}$. Hence the conserved classical charge is,
	\begin{equation}
	Q = i\int d^{3}x(\dot{\psi^{*}}\psi-\psi^{*}\dot{\psi}) = i\int d^{3}x (\pi \psi - \psi^{*} \pi^{*}),
	\end{equation}
	which after normal ordering becomes,
	\begin{equation}
	Q = \int \frac{d^{3}p}{(2\pi)^{3}}(c^{\dag}_{\vec{p}}c_{\vec{p}}-b^{\dag}_{\vec{p}}b_{\vec{p}}) = N_{c}-N_{b},
	\end{equation}
	where we have defined the $number \thinspace operator \thinspace N_{a\vec{p}}$ in the $n$-particle Hilbert space (Fock Space) as,
	\begin{equation}
	N_{a\vec{p}} = \int \frac{d^{3}p}{(2\pi)^{3}}a^{\dag}_{\vec{p}}a_{\vec{p}}.
	\end{equation}
	As a consequence of the Ward identities (appendix \ref{ward}), we have a conserved mean quantum charge $\langle Q \rangle = \langle N_{c}-N_{b}\rangle$ in the interacting theory.

	\section{Uniqueness of the Delta Distribution}
	\label{dirac}
	The constraint of a fixed energy necessarily means that there is a fixed mean energy (value $E$ say) and fixed higher order central moments (of value 0). A system with fixed energy classically corresponds to the micro canonical ensemble, this can be derived as follows. Instead of directly maximising the entropic functional,
	\begin{equation}
	S = -\int p_{\epsilon} \ln p_{\epsilon} - \alpha\bigg[\int p_{\epsilon} - 1\bigg]  - \sum^{\infty}_{i\geq 1}\lambda_{i}\bigg[\int p_{\epsilon}(\epsilon-\langle E \rangle)^{i} = 0 \bigg],
	\end{equation}
	(in which we have left out the $d\epsilon$ in the functional integrals to unclutter the notation). It can be shown that the delta distribution, satisfies the constraint of normalising to 1, has a mean of $E$ and has all higher order central moments being 0, i.e.
	\begin{equation}
	\int (E-E')^{n}\delta(E-E')dE' = (E-E)^{n} = 0.
	\end{equation}
	To prove that it is the only distribution that satisfies this infinite number of constraints, we use the definition of cumulants, $C_{n}(E)$, and characteristic functions, $\phi_{E}(k)$, to obtain,
	\begin{equation}
	\phi_{E}(k) = \langle e^{ikE}\rangle \equiv \int P_{E}(\epsilon)e^{ik\epsilon}d\epsilon =  \exp\bigg(\sum_{n=1}C_{n}(\epsilon)\frac{(ik)^{n}}{n!}\bigg),
	\end{equation}
	where the probability distribution $P_{E}$ is normalised to 1, $C_{1} = \langle E \rangle$ and $C_{2} = \langle E^{2} \rangle - \langle E \rangle^{2}$ etc. Setting all of the higher order cumulants other than $C_{1}$ to 0 yields,
	\begin{equation}
	\int P_{E}(\epsilon)e^{ik\epsilon}dx = \exp(\langle E \rangle ik).
	\end{equation}
	Differentiating the functional with respect to $e^{ikE}$ gives,
	\begin{equation}
	\begin{aligned}
	&\frac{\delta \phi_{E}(k)}{\delta e^{ik'E}} = P_{X}(x)\delta(k-k') = \delta(k-k')\delta(\langle E \rangle - E)\\ &\Rightarrow P_{X}(x) = \delta(\langle E \rangle -E).
	\end{aligned}
	\end{equation}

\noindent \textit{Bell Polynomials:} To show that this is equivalent to assuming that all the higher order central moments are 0, we write the $n$-th cumulant in terms of the central moments,
	\begin{equation}
	\begin{aligned}
	    &C_{n>1} = \sum_{k=1}^{n}(-1)^{k-1}(k-1)!B_{n,k}(0,\kappa_{2},\kappa_{n+1-k}) \\
	    &= \sum_{k=1}^{n}(-1)^{k-1}(k-1)!\sum\frac{n!}{\Pi_{a=1}^{n+1-k} j_{a}!}\bigg(\frac{\kappa_{a}}{n!}\bigg)^{j_{a}}, 	\end{aligned}
	\end{equation}
	\begin{equation}
	    \nonumber
	    \forall j_{a}\sum_{a=1}^{n+1-k} j_{a}=k,\sum_{a=1}^{n+1-k} aj_{a} = n.
	\end{equation}
	Where we have expanded the Bell polynomials. We note that the cumulant $C_{n}$ can be written as a combination of a multinomial power series in the central moments ($\kappa_{a}$), which we have specified to be 0 and hence the $C_{n} = 0$ for $n \geq 2$.   
	This concludes our derivation of the micro-canonical ensemble. Hence, for a time invariant Lagrangian, which by Noether's classical field theoretic result implies a fixed total energy, the only possible solution is the Dirac delta distribution.
	
\section{Central Limit Theorem}
\label{CLT}
Consider $S_{N} \equiv Y_{N}/N$ for large $N$, the distribution of $S_{N}$ converges towards a Gaussian with mean $\langle X \rangle$ and standard deviation $\sigma/\sqrt{N}$, provided higher order moments are finite.
\noindent \textbf{Proof:} via the characteristic function,
\begin{equation}
\begin{aligned}
&\phi_{S_{N}}(k) = \langle e^{ikS_{N}} \rangle \\
& = \langle e^{i\frac{k}{N}\sum_{j=1}^{N}X_{j}}\rangle =  \langle e^{i\frac{k}{N}X} \rangle^{N} = [\phi_{X}(k/N)]^{N},
\end{aligned}
\end{equation}
where we have used the properties of independence and identicality. In terms of cumulants,
\begin{equation}
\begin{aligned}
\exp \bigg(\sum_{m=0}^{\infty} \frac{(ik)^m}{m!}C_{m}(S_{N})\bigg) \\
= \exp \bigg(N\sum_{m=0}^{\infty} \frac{(ik/N)^m}{m!}C_{m}(S_{X})\bigg),
\end{aligned}
\end{equation}
thus
\begin{equation}
C_{M}(S_{N}) = C_{m}(X) N^{1-m}.
\end{equation}
Finite higher order moments all tend to $0$ as $N\rightarrow \infty$. As the Gaussian is uniquely defined by its mean and variance this becomes the limit distribution. This can be further seen by taking the inverse Fourier transform of the characteristic function and keeping the first two cumulants,
\begin{equation}
P_{S_{N}}(s) = \frac{1}{2\pi}\int_{\infty}^{\infty} e^{-iks}\phi_{S_{N}}(k)dk.
\end{equation}

\bibliographystyle{unsrt}
	\bibliography{sample}

\begin{thebibliography}{10}

\bibitem{Jaynes1967}
Edwin~T. Jaynes.
\newblock {\em Foundations of Probability Theory and Statistical Mechanics},
  pages 77--101.
\newblock Springer Berlin Heidelberg, Berlin, Heidelberg, 1967.

\bibitem{maxentreview}
Steve Press\'e, Kingshuk Ghosh, Julian Lee, and Ken~A. Dill.
\newblock Principles of maximum entropy and maximum caliber in statistical
  physics.
\newblock {\em Rev. Mod. Phys.}, 85:1115--1141, Jul 2013.

\bibitem{inftheoryjaynes}
E.~T. Jaynes.
\newblock Information theory and statistical mechanics.
\newblock {\em Phys. Rev.}, 106:620--630, May 1957.

\bibitem{shore1980axiomatic}
John Shore and Rodney Johnson.
\newblock Axiomatic derivation of the principle of maximum entropy and the
  principle of minimum cross-entropy.
\newblock {\em IEEE Transactions on information theory}, 26(1):26--37, 1980.

\bibitem{giffin}
Adom Giffin, Carlo Cafaro, and Sean~Alan Ali.
\newblock Application of the maximum relative entropy method to the physics of
  ferromagnetic materials.
\newblock {\em Physica A: Statistical Mechanics and its Applications}, 455:11
  -- 26, 2016.

\bibitem{nerioptions}
Cassio Neri and Lorenz Schneider.
\newblock Maximum entropy distributions inferred from option portfolios on an
  asset.
\newblock {\em Finance and Stochastics}, 16(2):293--318, 2012.

\bibitem{entropybuchen}
Peter~W Buchen and Michael Kelly.
\newblock The maximum entropy distribution of an asset inferred from option
  prices.
\newblock {\em Journal of Financial and Quantitative Analysis},
  31(01):143--159, 1996.

\bibitem{Gonzalez2014}
Diego Gonz{\'a}lez, Sergio Davis, and Gonzalo Guti{\'e}rrez.
\newblock Newtonian dynamics from the principle of maximum caliber.
\newblock {\em Foundations of Physics}, 44(9):923--931, 2014.

\bibitem{caticha2012entropic}
A~Caticha.
\newblock Entropic inference and the foundations of physics (monograph
  commissioned by the 11th brazilian meeting on {B}ayesian
  statistics--ebeb-2012, 2012.

\bibitem{brandeis}
R.~D. Rosenkrantz.
\newblock {\em Brandeis Lectures (1963)}, pages 39--76.
\newblock Springer Netherlands, Dordrecht, 1989.

\bibitem{julianandpresse}
Julian Lee and Steve Press\'e.
\newblock Microcanonical origin of the maximum entropy principle for open
  systems.
\newblock {\em Phys. Rev. E}, 86:041126, Oct 2012.

\bibitem{chandler}
David Chandler.
\newblock {\em Introduction to Modern Statistical Mechanics}.
\newblock Oxford University Press, 1987.

\bibitem{lee}
Julian Lee.
\newblock Microcanonical origin of the maximum entropy principle for open
  systems.
\newblock {\em arxiv}, 2012.

\bibitem{landau_statphys}
L.~D. Landau, Lifshitz~E. M., and Pitaevskiĭ~L. P.
\newblock {\em Statistical physics}.
\newblock Elsevier, 2012.

\bibitem{qmhermitian}
Alessio Benavoli, Alessandro Facchini, and Marco Zaffalon.
\newblock Quantum mechanics: The {B}ayesian theory generalized to the space of
  {H}ermitian matrices.
\newblock {\em Physical Review A}, 94(4), Oct 2016.

\bibitem{singlemoleculedna}
Carlos Bustamante, Zev Bryant, and Steven~B. Smith.
\newblock Ten years of tension: single-molecule {DNA} mechanics.
\newblock {\em Nature}, 421(6921):423–427, 2003.

\bibitem{latticetsallis}
P.~Douglas, S.~Bergamini, and F.~Renzoni.
\newblock Tunable {T}sallis distributions in dissipative optical lattices.
\newblock {\em Physical Review Letters}, 96(11), 2006.

\bibitem{optiontsallis}
Lisa Borland.
\newblock Option pricing formulas based on a non-{G}aussian stock price model.
\newblock {\em Physical Review Letters}, 89(9), Jul 2002.

\bibitem{qreview}
S.~Picoli Jr., R.~S. Mendes, L.~C. Malacarne, and R.~P.~B. Santos.
\newblock q-distributions in complex systems: a brief review.
\newblock {\em Brazilian Journal of Physics}, 39(2a):468–474, 2009.

\bibitem{cox}
R.~T. Cox.
\newblock Probability, frequency and reasonable expectation.
\newblock {\em American Journal of Physics}, 14(1):1--13, 1946.

\bibitem{stein1972}
Charles Stein.
\newblock A bound for the error in the normal approximation to the distribution
  of a sum of dependent random variables.
\newblock In {\em Proceedings of the Sixth Berkeley Symposium on Mathematical
  Statistics and Probability, Volume 2: Probability Theory}, pages 583--602,
  Berkeley, Calif., 1972. University of California Press.

\bibitem{convergenceclt}
A.~N. Tikhomirov.
\newblock On the convergence rate in the central limit theorem for weakly
  dependent random variables.
\newblock {\em Theory of Probability 'l\&' Its Applications}, 25(4):790–809,
  1981.

\bibitem{tsallis_2009}
Constantino Tsallis.
\newblock {\em Introduction to nonextensive statistical mechanics: approaching
  a complex world}.
\newblock Springer, 2009.

\bibitem{whatistsallisq}
G.~Wilk and Z.~Włodarczyk.
\newblock Interpretation of the nonextensivity parameterqin some applications
  of {T}sallis statistics and {L}\'{e}vy distributions.
\newblock {\em Physical Review Letters}, 84(13):2770–2773, 2000.

\bibitem{beck_2001}
Christian Beck.
\newblock Dynamical foundations of nonextensive statistical mechanics.
\newblock {\em Physical Review Letters}, 87(18), Oct 2001.

\bibitem{tsalliscritiquethermal}
Michael Nauenberg.
\newblock Critique of $q$-entropy for thermal statistics.
\newblock {\em Physical Review E}, 67(3), 2003.

\bibitem{tsalliscritiquedata}
Steve Pressé, Kingshuk Ghosh, Julian Lee, and Ken~A. Dill.
\newblock Nonadditive entropies yield probability distributions with biases not
  warranted by the data.
\newblock {\em Physical Review Letters}, 111(18), Jan 2013.

\bibitem{boundedqleq1}
Alberto d'Onforio.
\newblock Bounded noises in physics, biology, and engineering.
\newblock {\em Modeling and Simulation in Science, Engineering and Technology},
  2013.

\end{thebibliography}
\end{document}